\documentclass[sigplan]{acmart}

\usepackage{booktabs} \usepackage{opera}
\usepackage{amsthm}
\usepackage{hyperref}
\usepackage{pifont}
\usepackage[framemethod=tikz]{mdframed}
\setcopyright{none}

\acmDOI{10.475/123_4}

\acmISBN{123-4567-24-567/08/06}

\acmConference[TaPP'18]{USENIX Workshop on Theory and Practice of Provenance}{July 2018}{London, United Kingdom}
\acmYear{2018}
\copyrightyear{2018}

\usepackage{titlesec}
\titlespacing*{\section}{0pt}{0.2\baselineskip}{0.2\baselineskip}

\begin{document}
\title{Provenance-based Intrusion Detection: Opportunities and Challenges}

\author{Xueyuan Han}
\affiliation{  \institution{Harvard University}
}

\author{Thomas Pasquier}
\orcid{0000-0001-6876-1306}
\affiliation{  \institution{University of Cambridge}
}

\author{Margo Seltzer}
\affiliation{  \institution{Harvard University}
}

\renewcommand{\shortauthors}{Han et al.}

\renewcommand{\figureautorefname}{Fig.}
\renewcommand{\tableautorefname}{Table}
\renewcommand{\sectionautorefname}{\S}
\renewcommand{\subsectionautorefname}{\S}
\providecommand*{\lstlistingautorefname}{Listing}

\begin{abstract}
\makeatletter{}Intrusion detection is an arms race;
attackers evade intrusion detection systems by developing new attack vectors to sidestep known defense mechanisms.
Provenance provides a detailed, structured history of the interactions of digital objects within a system.
It is ideal for intrusion detection, because it offers a \textit{holistic}, attack-vector-agnostic view of system execution.
As such,
provenance graph analysis fundamentally strengthens detection robustness. 
We discuss the opportunities and challenges associated with provenance-based intrusion detection
and provide insights based on our experience building such systems.
 
\end{abstract}

\maketitle

\section{Introduction}
\label{sec:introduction}
\makeatletter{}System security continues to be an arms race between intruders and defenders.
In this arms race,
attackers adapt in response to defense mechanisms and \emph{always} win.
Defeating attackers requires
rethinking traditional defeat- and exploit-based mitigation techniques,
which lack complete security coverage~\cite{pincus2004mitigations}.
We propose taking
a holistic, attack-vector-agnostic view of system execution.

\emph{We claim that provenance is the ideal data to use for such a task
and that provenance graph-based analysis is the ultimate means towards
achieving complete security coverage.}
Provenance refers to meta-data describing how digital objects came to be
in their current state.
It provides a complete, structured view of what happened on the system~\cite{bates2015trustworthy} by
presenting complex dependencies and causality relationships between digital objects as a directed acyclic graph (DAG).
As such,
it is well suited for intrusion detection.
An intrusion manifests in anomalous interdependencies among data objects
that deviate from those found in non-malicious execution.
In fact,
in attack causality analysis~\cite{king2005enriching},
provenance has long been used to explain intrusions (~\autoref{sec:oc}).

Provenance graph analysis strengthens adversarial robustness, because
the graphs exhibit long-range correlations and dependencies
allowing for causal reasoning about intrusions~\cite{akoglu2015graph}
(\autoref{sec:applicability}).
Such causal reasoning enables detection of sophisticated attacks, such as 
network attacks, that remain undetected for long periods of time.
In prior work~\cite{han2017frappuccino}, we reduced a host-based
intrusion detection problem to a graph-based anomaly detection problem,
in which graph analysis identified structured execution
traces that represented an intrusion.
However,
intrusion detection on provenance graphs requires analyzing dynamic, attributed, streaming graphs, which are rarely studied in the literature.
Given the development of fine-grained, whole-system provenance
capture systems~\cite{pasquier2017practical},
this task becomes even more challening
as the graphs rapidly become extraordinarily large~\cite{bates2015trustworthy}.
However, we can use
domain-specific knowledge of provenance graphs to simplify the challenge of
identifying anomalies, making it an easier problem than general-purpose graph
analysis suggests.
For example, since execution history is immutable,
we can assume that provenance graphs only increase in size (i.e., there are
never deletions).
This property allows us to incrementally and progressively reason about
causality without needing to look backwards.

\section{Applicability}
\label{sec:applicability}
\makeatletter{}\newmdtheoremenv[
hidealllines=true,
leftline=true,
innertopmargin=0pt,
innerbottommargin=0pt,
linewidth=4pt,
linecolor=gray!40,
innerrightmargin=0pt,
]{definitiona}{Definition}

Data provenance has seen use in areas such as databases and computational sciences.
While it now also appears as part of
real-time security analysis~\cite{bates2015trustworthy},
most approaches are variations of dynamic taint analysis of provenance data.
While simple and effective on their own merits,
they are limited to constraining information flows within a system
(\eg data loss prevention, access control, and regulatory compliance);
little work has been done to detect intrusions from outside the system~\cite{han2017frappuccino, pasquier2017practical}.

Host-based anomaly detection systems define some baseline normal behavior
and then classify as abnormal any behavior that significantly deviates from the baseline.
The approach is predicated on the assumption that intrusions are highly correlated to abnormal behavior.
Many existing systems use unstructured collections of multidimensional data (\eg audit logs)
to detect outlying points in a high-dimensional feature space,
formulating intrusion detection as point-based outlier detection 
to leverage various learning and data mining techniques.
Provenance, however, 
is structured graph data that represents relationships between a digital item (\ie data entity),
a transformation on that item (\ie activity),
and agents (\ie persons and organizations) associated with the item and the transformation. 
Hence,
unlike the prior work,
we formulate the host-based intrusion detection problem as
a graph-based anomaly detection problem defined as
follows~\cite{akoglu2015graph}:
\begin{definitiona}
\noindent{The graph-based intrusion detection problem is to identify components of the graph that are significantly different from those in a learned model of the graph.}
\end{definitiona}

Using a provenance graph-based approach to intrusion detection is suitable for various reasons:

\noindent{{\tiny\ding{108}}}~\noindemph{Provenance captures complete access to security-sensitive kernel objects:}
State-of-the-art provenance whole-system capture systems leverage the Linux Security Module (LSM) interface
to record provenance for every security-related interaction, rather than intercepting system calls.
They can be extended to verifiably monitor all information flows in a system~\cite{georget2017verifying}.

\noindent{{\tiny\ding{108}}}~\noindemph{Provenance makes explicit the relationships among objects:} 
One powerful feature of provenance is its native graphical representation
to show system execution as interactions between data objects.
However, 
such interdependencies are innate to every execution trace,
even in seemingly unstructured audit data from logging systems such as \texttt{auditd}.
In fact,
there exist frameworks that reconstruct graph-based provenance from flat audit data to allow for reasoning about system execution~\cite{gehani2012spade}. 
However,
this post hoc approach comes with a caveat: it is harder to ensure completeness or correctness of the graph built from flat audit data~\cite{pohly2012hi}.

\noindent{{\tiny\ding{108}}}~\noindemph{Intrusions result from unexpected interactions:}
The entry point to a victim system may be a single, isolated event,
but its effects must propagate for an intrusion to be fruitful to an attacker.
For example,
consider an insider attacker who wishes to steal sensitive information from a data server under his control.
He first installs a malicious BASH script 
that discovers and collects all documents (\ie a single entry point to the server).
However,
to successfully steal the information, 
he needs to either transfer it to a foreign machine or write it to an external storage device.
The key to detecting the data leak is to connect the collection of the data to the transmission of the data,
which in a provenance graph is clearly represented as a chain of dependencies between processes, files, and sockets.

\noindent{{\tiny\ding{108}}}~\noindemph{Graph representation improves robustness:}
graphs are generally more adversarially robust,
\ie it is harder for an attacker to camouflage her behavior to fit into the reference graph structures~\cite{akoglu2015graph}.
In fact,
we claim that \emph{the provenance graph of an intrusion must differ from that of a valid execution when we use an LSM-based whole-system provenance capture system.}
As LSM places hooks on any execution path that generates an information flow~\cite{akoglu2015graph},
if the capture system records provenance on every such path,
violations of security policies will be evident from the provenance graph.
Moreover,
the attacker must also have the knowledge of the substructures that are referenced by the IDS,
which alone requires significant effort.
For example,
the attacker from the previous example may evade detection if each step is allowed when performed in isolation.
An advanced attacker can even fake the IP address of the foreign machine.
However,
when considering the chain of actions as a whole (\ie an abnormal graph substructure),
we can identify the intrusion.

\section{Opportunities and Challenges}
\label{sec:oc}
\makeatletter{}Analyzing dynamic, attributed graphs is difficult.
Graph anomaly detection in this setting requires detecting changes over time,
which in turn requires a formal notion of similarity defined specifically for the target domain~\cite{akoglu2015graph}.
With attributed vertices and edges, 
changes can occur both structurally and in labels.
Provenance graphs further complicate the matter as each vertex and edge usually has a set of attributes
(instead of a single \texttt{type} attribute),
and the number of attributes varies depending on the type of the vertex/edge.
To enable online intrusion detection,
one also receives the provenance graph in a streaming fashion
and must perform the analysis in realtime.
However,
provenance graphs are acyclic, thus having a topological ordering that simplifies computation.
Events therefore can be partially ordered as they are streamed for analysis~\cite{pasquier2017practical}.
We can then efficiently reason over the vast amount of information contained in vertex and edge labels,
which, combined with structural information,
reflects various aspects of system execution.
In the following sections, 
we discuss the main opportunities and challenges associated with provenance-based intrusion detection.

\subsection{Opportunities}
\label{sec:opportunities}
\noindgras{Opportunity 1: Provenance graph structures and labels encode the complete, historical context of system execution.}
A useful intrusion detection system learns detailed normal behavior from the past.
Given flat audit data with no completeness guarantee,
an IDS is limited by the data recorded in the audit logs.
It is also difficult to obtain higher-order dependencies~\cite{ye2000markov}. 
In some cases,
the type of information it learns from is determined empirically by the attack vectors it is designed to detect.
Such an ad hoc approach ultimately leads to the arms race described in~\autoref{sec:introduction}.
In contrast,
whole-system provenance provides a complete view of information flow
that natively reflects higher-order correlations and long-range dependencies.
Its graph structure also allows for graph-based analysis. 
We illustrate its benefits by describing the following principles
that provenance analysis embodies.

\noindent{{\tiny\ding{108}}}~\noindemph{Principle 1: Identify semantically meaningful substructures.}
Provenance graphs can become large,
obfuscating important events that require special attention.
Complex system interactions within a task and between tasks further cloud understanding.
Therefore,
it is important to identify substructures/subgraphs that are semantically coherent (\eg describing a single task within a program).
Macko \etal~\cite{macko2013local} developed two centrality metrics to perform local clustering on provenance graphs for task separation.
Generic metrics used to discover communities are also applicable, albeit expensive in certain cases.
Significant changes in those structures usually imply intrusions.
For example,
most control-data attacks alter the control flow of a program to execute injected malicious code.
They typically start a new shell with the privilege of the victim process~\cite{chen2005non},
which inevitably introduces unexpected vertices and edges in the provenance graph.
Akoglu \etal~\cite{akoglu2015graph} summarized various distance measures to detect structural anomalies in dynamic graphs.

\noindent{{\tiny\ding{108}}}~\noindemph{Principle 2: Incorporate time.}
The rate of provenance event creation is proportional to kernel object access rate.
As each access to security-sensitive kernel object results in (at least) one edge in the graph, 
provenance graphs reflect this rate through the number of vertices and/or edges per unit of time.
Although some 
benign workloads exhibit a high rate of provenance generation (\eg building a kernel~\cite{pasquier2017practical}),
bursts of intense provenance generation frequently indicate an attack.
For example,
attackers exploit race conditions to deploy Time-of-Check-to-Time-of-Use (TOCTOU) attacks.
The fairly recent Dirty COW attack(CVE-2016-5195),
in which the Linux kernel's memory subsystem incorrectly handled
copy-on-write (COW), granting write access to private read-only memory mappings,
used two threads simultaneously bombarding the system with \texttt{madvise}
and \texttt{write} system calls. These calls produce elements of the provenance
graph at a rate rarely observed during normal behavior.

\noindent{{\tiny\ding{108}}}~\noindemph{Principle 3: Keep history in mind.}
Advanced persistent threat (APT) attacks are usually a set of continuous,
long-running processes that permeate the victim system.
Noticing such attacks requires a holistic understanding of system execution
starting from its initialization.
In fact,
any intrusion that requires retrospective analysis on previously processed
portion of the graph 
can be discovered only if the detection system ``remembers'' history.
However,
the sheer volume of provenance data renders any attempt at a complete review impractical.
One way to mitigate this needle-in-a-haystack problem is to incrementally build a concise yet comprehensive model
that memorizes the historical context of the graph.
For example, 
Lemay \etal~\cite{lemay2017automated} designed regular grammars for provenance
DAGs to succinctly summarize the graph structure.

\noindgras{Opportunity 2: Provenance graphs are topologically and partially ordered.}
This property follows naturally from the fact that provenance graphs are DAGs
and that they truthfully reflect the causal relationships of events that occurred on the system.
We took advantage of this property and designed a real-time provenance analysis framework to enable semantically rich security services~\cite{han2017frappuccino}.
In particular,
a vertex-centric graph framework facilitates provenance graph analysis with its correctness guaranteed by the two partial ordering properties:
1) once an outgoing edge to a vertex arrives,
we know that we have observed all incoming edges to that vertex;
2) we receive all edges and vertices along a path in order.

\noindgras{Opportunity 3: Provenance graphs enrich attack attribution and sense-making.}
Attribution is an important feature that allows system administrators to quickly understand the source of an intrusion
so that they can remedy the issue in a timely fashion and effectively control the damage.
Many intrusion detection systems suffer from a high false positive rate.
Attribution helps administrators quickly reject false positive alarms,
effectively making the IDS more usable.
Provenance graphs are causality graphs that naturally allow for sense-making,
providing a causal chain of events for reasoning.
For example,
King \etal~\cite{king2003backtracking} designed a system that structures OS-level audit logs to automatically identify sequences of steps that occurred in an intrusion, starting from a single detection point.

\subsection{Challenges}
\label{sec:challenges}
\noindgras{Challenge 1: It is difficult to obtain a good graph summary.}
From Opportunity 1 (\autoref{sec:opportunities}), 
we see that a good graph summary should at least adhere to all the principles discussed.
For principles that do not consider graph structures,
we can learn trends via applications of machine learning or empirically,
e.g., by finding and setting a threshold.
However, 
the streaming nature of provenance data for online intrusion detection makes graph analysis challenging.
One approach is to segment the graph using a time window,
though one needs to determine an appropriate window size.

\noindgras{Challenge 2: Online intrusion detection requires efficient computation.}
Even with the framework described in \autoref{sec:opportunities},
the computation itself (\eg to generate a good graph summary) must be efficient
enough to detect an intrusion before it wreaks havoc on the system.
Many intrusion detection systems require training on known datasets,
which is often performed offline~\cite{axelsson2000intrusion}.
Efficiency therefore is usually a primary concern during deployment.
Complicated graph algorithms, such as subgraph isomorphism,
are often NP-complete and are suitable only for small graphs.
Machine learning and data mining approaches on graphs,
\eg graph kernels,
offer alternatives with polynomial or even linear time complexity.

\noindgras{Challenge 3: The complexity of the system makes provenance graphs difficult to understand.}
There exists a trade-off between the completeness of provenance and the succinctness of the resulting graph.
With whole-system provenance capture,
this trade-off becomes even clearer as a large number of underlying system dependencies are captured.
For example,
Liu \etal~\cite{liutowards} showed that a simple \texttt{sshd} command can trigger a massive number of Linux commands that are used to update Linux environment variables,
which results in a large provenance subgraph describing these activities.
However,
they also proposed an algorithm that takes into account factors, such as rareness and dataflow termination,
to determine the priority of events during backward and forward tracking of a provenance graph.

\section{Experience}
\label{sec:experience}
\makeatletter{}In prior work~\cite{han2017frappuccino},
we presented a provenance-based intrusion detection system.
As we refined our system~\cite{pasquier2017practical},
we identified idiosyncrasies that differentiate intrustion detection
via provenance and via audit logs.
In addition to the properties already discussed,
provenance captures
interactions across applications that are invaluable
in intrusion detection.

Based on our prior experience, we identify the following keys to
provenance-based intrusion detection:

\noindent{{\tiny\ding{108}}}~\noindemph{Understand the provenance capture mechanism and the graph it produces:}
It is important to understand what information is captured, how it is captured, and at what level of granularity.
These all affect graph interpretation.
For example, 
we have worked with capture systems that record both
thread-level details~\cite{pasquier2017practical} and
process-only details~\cite{gehani2012spade}.
They have fundamentally different underlying capture mechanisms,
and therefore,
we need to make different assumptions about the provenance graphs they generate,
even when they are capturing provenance of the same system execution.
More importantly, 
we need to make correct assumptions,
which is fundamental to the correctness of any provenance graph analysis.
Consequently,
it is essential to specify the formalization of the graphs from different capture mechanisms, not to generalize.

Sometimes,
existing provenance capture systems may not fulfill the needs of an IDS;
jointly developing a provenance capture system and a provenance-based IDS is most likely to improve the performance of both systems.

\noindent{{\tiny\ding{108}}}~\noindemph{Build datasets to benchmark IDSes:}
The lack of labeled datasets is a serious obstacle to work in this area.
As provenance capture mechanisms evolve,
a plug-and-play system that can automatically rerun experiments
is valuable.
We use Vagrant to generate experimental
data in a virtual environment~\cite{provdata}. 
However, 
labeling datasets is tricky~\cite{maggi2010detecting}.
One cannot simply label an entire provenance graph as an ``intrusion'',
since an IDS could mistakenly interpret a benign subgraph as
an intrusion entry point.
On the other hand,
a provenance graph of seemingly normal system execution might contain
unexpected execution errors,
which, though not part of an intrusion,
still deviate from specified normal behavior.
This difficulty leads to misleading comparison metrics,
such as precision, recall, and F-measure.
Benchmarking IDSes remains an important open problem.

\section{Conclusion}
\label{sec:conclusion}
\makeatletter{}We propose to realize robust, attack-vector-agnostic intrusion detection through analysis on provenance graphs
and identify opportunities and challenges specific to whole-system,
provenance-based intrusion detection.
While the concept of OS-level provenance is almost a decade old,
formalization and theoretical studies of its graphs have not yet materialized.
Applying whole-system provenance to 
intrusion detection~\cite{han2017frappuccino}
requires a formal understanding of provenance.
We invite fellow researchers in both theory and provenance communities to
continue this exploration with us.

\bibliographystyle{ACM-Reference-Format}
\bibliography{bibliography}

\end{document}